\RequirePackage{fix-cm}
\documentclass[twocolumn]{svjour3}          % twocolumn
\smartqed  % flush right qed marks, e.g. at end of proof
\usepackage{graphicx,amsmath,amssymb,color}
\journalname{Shock Waves}
\graphicspath{{../data/}}

\newcommand{\be}{\begin{equation}}
\newcommand{\ee}{\end{equation}}
\newcommand{\bea}{\begin{eqnarray}}
\newcommand{\eea}{\end{eqnarray}}

\newcommand{\dv}[2]{\frac{d#1}{d#2}}

\newcommand\Pran{\mbox{\textit{Pr}}} 

\newcommand\cm{{\rm\,cm}}
\newcommand\gcm{{\rm\,g\,cm^{-3}}}
\newcommand\kev{{\rm\,keV}}

\begin{document}

\title{{\color{black}Three temperature} plasma shock solutions with gray radiation diffusion%\thanks{Grants or other notes
}


\author{Bryan M. Johnson \and Richard I. Klein %etc.
}


\institute{Bryan M. Johnson \at
              Lawrence Livermore National Laboratory, 7000 East Ave., Livermore, CA 94550 \\
              \email{johnson359@llnl.gov} \\
              \\
              Richard I. Klein \at
              Lawrence Livermore National Laboratory, 7000 East Ave., Livermore, CA 94550,
              {\color{black}and University of California at Berkeley, Department of Astronomy} \\
}

\date{Received: 8 October 2015 / Accepted: 17 March 2016}

\maketitle

\begin{abstract}
The effects of radiation on the structure of shocks in a fully-ionized plasma are investigated by solving the steady-state fluid equations for ions, electrons, and radiation. The electrons and ions are assumed to have the same bulk velocity but separate temperatures, and the radiation is modeled with the gray-diffusion approximation. Both electron and ion conduction are included, as well as ion viscosity. When the material is optically thin, three-temperature behavior occurs. When the diffusive flux of radiation is important but radiation pressure is not, two-temperature behavior occurs, with the electrons strongly coupled to the radiation. Since the radiation heats the electrons on length scales that are much longer than the electron-ion Coulomb coupling length scale, these solutions resemble radiative shock solutions rather than plasma shock solutions that neglect radiation. When radiation pressure is important, all three components are strongly coupled. Results with constant values for the transport and coupling coefficients are compared to a full numerical simulation with a good match between the two, demonstrating that steady shock solutions constitute a straightforward and comprehensive verification test methodology for multi-physics numerical algorithms.
\keywords{Plasma shocks \and Radiative shocks \and Code verification}
\end{abstract}

\section{\label{INTRO}Introduction}

The spatial structure of a shock propagating through an ionized gas is a classic problem in plasma physics \cite{zr67}. In addition to its myriad physical applications, the problem is an excellent test-bed for studying multiple coupled physical effects, providing both physical insight and a framework for multi-physics code verification. Previous studies have ignored the effects of radiation on plasma shocks \cite{zr67,jp64,mwl11}; radiative effects have only been considered for a gas {\color{black}with a two-temperature (material plus radiation) structure} \cite{zr67,mm84,le08}.

The purpose of this work is to explore the impact of radiation on plasma shocks by solving for the shock structure in a three-temperature system: ions, electrons, and radiation. For simplicity, the ions and electrons will be assumed to have the same bulk velocity but separate internal energies, and the radiation will be treated in the gray diffusion approximation. The steady-state equations to be solved are given in \S\ref{EQUATIONS}, and the methodology employed in solving them is described in \S\ref{METHOD}, along with a discussion of the difficulties associated with applying standard methods to this system. Results are given in \S\ref{RESULTS}, followed by a discussion in \S\ref{DISCUSSION}.

\section{\label{EQUATIONS}Steady-State Equations}

Upon integrating the continuity equation to obtain a constant mass flux $m_0 = \rho v$, where $\rho$ is the mass density and $v$ is the bulk flow velocity in the frame of the shock, the steady-state equations to be solved are
\be\label{MOM}
m_0 \dv{v}{x} + \dv{p}{x} = -\dv{F_{v}}{x},
\ee
\be\label{EE}
m_0\dv{e_e}{x} + p_e\dv{v}{x} = -S_{ei} - \dv{F_e}{x} - S_{er},
\ee
\be\label{EI}
m_0\dv{e_i}{x} + p_i\dv{v}{x} = S_{ei} - \dv{F_i}{x} + \frac{4\mu_i}{3}\left(\dv{v}{x}\right)^2,
\ee
\be\label{ER}
m_0 \dv{e_r}{x} + p_r \dv{v}{x} = - \dv{F_r}{x} + S_{er},
\ee
where $p_\alpha$, $T_\alpha$, $e_\alpha$ denote pressure, temperature and specific energy (with $\alpha = e$, $i$ and $r$ for electron, ion and radiation quantities, respectively), $p = p_e + p_i + p_r$ is the total pressure of the three species, $F_{v} = -(4\mu_i/3)dv/dx$ is the ion viscous momentum flux ($\mu_i$ is the ion viscosity), $F_\alpha = -\kappa_\alpha dT_\alpha/dx$ is the species heat flux ($\kappa_\alpha$ is the conductivity), and $S_{ei}$ and $S_{er}$ are coupling terms. The radiation energy equation (\ref{ER}) is in the mixed frame of radiation hydrodynamics \cite{mm84,mk82} and has been expressed in a form that parallels the material energy equations ($e_r = a_r T_r^4/\rho$ is the radiation energy per unit mass of material, where $a_r$ is the radiation constant).

The radiation couples to the material energy through the electrons:
\[
S_{er} = c \chi_a a_r\left(T_e^4 - T_r^4\right),
\]
where $\chi_a$ is the absorption opacity and $c$ is the speed of light, and the electron and ion internal energies are coupled via
\[
S_{ei} = \Gamma_{ei}\left(T_e - T_i\right),
\]
where $\Gamma_{ei}$ is the electron-ion energy coupling parameter. The radiative conductivity is
\[
\kappa_r = \frac{4 c a_r T_r^3}{3\chi_t},
\]
where $\chi_t = \chi_a + \chi_s$ is the total opacity and $\chi_s$ is the scattering opacity.

The electrons and ions are assumed to obey ideal-gas equations of state, $p_{e,i} = (\gamma - 1) \rho e_{e,i}$ ($\gamma$ is the adiabatic index), with $e_{e,i} = C_{ve,i} T_{e,i}$, whereas the radiation obeys the equation of state $p_r = (1/3)\rho e_r$ {\color{black}(this is strictly true only in the optically thick limit)}. The material specific heats are taken to be $C_{ve} = C_{vi} = C_v/2$, where
\[
C_v \equiv \frac{k_B}{m_i\left(\gamma - 1\right)}
\]
is the total material specific heat, $k_B$ is Boltzmann's constant and $m_i$ is the ion mass (with these definitions, $e_e$ is the electron energy per ion mass).\footnote{{\color{black}We have chosen $C_{ve} = C_{vi}$ for simplicity; this is not a general property of plasmas, nor is it necessary for our solution methodology. It is a valid physical assumption for a hydrogen plasma.}}

The system of equations (\ref{MOM})--(\ref{ER}) admits two integrals, one from (\ref{MOM}) and one from the total energy equation obtained by summing {\color{black}$v$ times (\ref{MOM}) and} (\ref{EE})--(\ref{ER}):
\be\label{INT1}
v + \frac{p}{m_0} + \frac{F_{v}}{m_0} = v_0 + \frac{p_0}{m_0} \equiv c_1,
\ee
\be\label{INT2}
\frac{v^2}{2} + h + \frac{F}{m_0} = \frac{v_0^2}{2} + h_0 \equiv c_2,
\ee
where $h = \gamma e_e + \gamma e_i + (4/3)e_r$ is the total enthalpy of the three species, $F = F_e + F_i + F_r + v F_{v}$ is the total dissipative flux, a zero subscript denotes an ambient fluid quantity and the dissipative fluxes have been taken to be zero at the boundaries. 

Solving equations (\ref{INT1}) and (\ref{INT2}) with $F = F_{v} = 0$ yields the shock jump conditions. The hydrodynamic shock jump conditions are specified by a single parameter, the Mach number ${\cal M}_0 \equiv v_0/a_0$ of the shock, where $a_0 = \sqrt{\gamma k_B T_0/m_i}$ is the ambient material sound speed. Radiation adds two additional parameters, which we will take to be $\rho_0$ and $T_0$, the ambient density and temperature. One can alternatively express the two additional parameters as in \cite{le08}:
\[
{\cal C}_0 \equiv \frac{c}{a_0}, \;\; {\cal P}_0 \equiv \frac{a_r T_0^4}{\rho_0 a_0^2},
\]
where the latter parameter gives a measure of the importance of radiation pressure relative to material pressure. 

\section{\label{METHOD}Methodology}

Extracting shock solutions from equations (\ref{MOM})--(\ref{ER}) is not trivial \cite{nem60,and63,gro65}. The primary reason for this is that other solutions exist in addition to the shock solution, and the non-shock solutions can be a stronger attractor (in the steady-state domain) than the solution of interest. Examples of non-shock solutions can be derived from a simplified version of our equation set, and we describe some of these in the Appendix \cite{bec22,joh13,joh14}. {\color{black}Most previous work has employed a shooting method for a system of two differential equations; the stability properties of such a system are straightforward to analyze, and one can usually find a stable method of obtaining a shock solution.

The only study we know of that employs shooting with more than two equations is reference \cite{juk57}. In that study, a system of three equations is solved with a form of shooting, using a linear analysis near the end points as a guide. Analyzing the general stability properties of our system of equations is beyond the scope of this paper, but in principle one could proceed in a manner similar to \cite{juk57} with an arbitrary number of equations. Each spatial derivative increases the order of the linear eigenvalue problem that must be solved, however, and with it the complexity of the stability analysis. The feasibility of such a method may also depend upon the particular landscape in parameter space associated with a given system of equations.}

To avoid the difficulties associated with shooting methods, we have chosen instead to solve equations (\ref{MOM})--(\ref{ER}) with relaxation \cite{pre92}. Rather than integrating from the endpoints, a guess for the entire solution is initialized on a grid, and the solver attempts to iterate to convergence. The primary challenge of this method is coming up with a good initial guess. Once a solution is obtained, it is fairly straightforward to step through parameter space to obtain other solutions, although even this must be done with care. The Appendix discusses the initial guess that was used and gives some additional pointers for obtaining solutions with this method. Due to the inclusion of viscosity, the solutions are continuous, even when they contain an inner viscous layer.

A summary of the basic solution procedure is as follows (see the Appendix for details):
\begin{enumerate}
\item Calculate the shock jump conditions from expression (\ref{INT1}) and (\ref{INT2}) with a root finding algorithm \cite{pre92}.
\item Initialize temperatures and fluxes with the analytical solution given by expressions (\ref{TI})--(\ref{FTI}) in the Appendix.
\item Obtain an initial solution with constant coefficients using a relaxation algorithm \cite{pre92}. The opacity and electron-ion coupling coefficient used here should be sufficiently large that all three temperatures are well coupled, and viscosity should dominate conductivity.
\item Slowly transition from a solution with constant coefficients to a solution with physical coefficients (or to a solution with different values for the constant coefficients), solving the relaxation algorithm at each step of the transition.
\end{enumerate}

\section{\label{RESULTS}Results}

\begin{figure}
\includegraphics[scale=0.45]{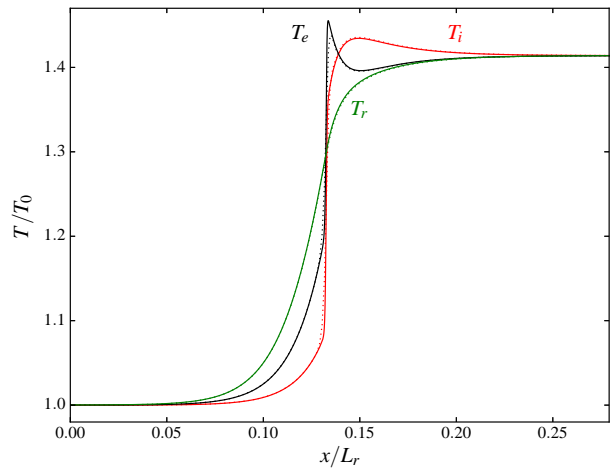}
\caption{\label{F1}Electron (\emph{black}), ion (\emph{red}) and radiation (\emph{green}) temperature profiles for a low Mach number radiative plasma shock with constant coefficients. Dotted lines are \texttt{Kull} results.}
\end{figure}

\begin{figure}
\includegraphics[scale=0.45]{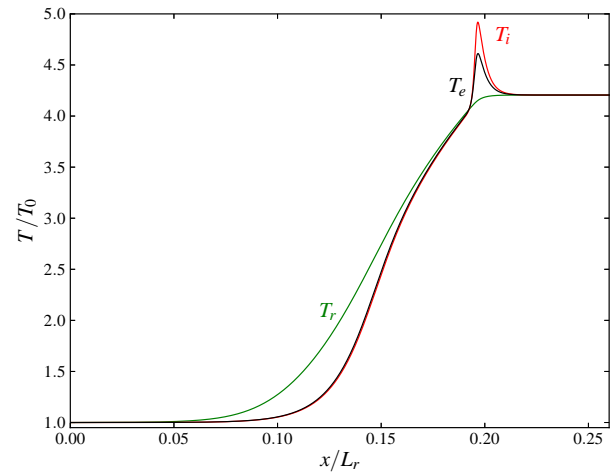}
\caption{\label{F2}Electron (\emph{black}), ion (\emph{red}) and radiation (\emph{green}) temperature profiles for a high Mach number radiative plasma shock with constant coefficients.}
\end{figure}

We first show results with constant coefficients to demonstrate our approach and to make a comparison with a hydrodynamics code. Figures~\ref{F1} and \ref{F2} show results for both a low- and high-Mach number radiative plasma shock; the shocks  propagate from right to left in the lab frame. The common parameters used here were $\gamma = 5/3$, $C_{ve} = C_{vi} = 1$, $\rho_0 = 40 \gcm$, $T_0 = 2 \kev$ (corresponding to ${\cal C}_0 = 142.204$, ${\cal P}_0 = 0.001234815$), $\kappa_e = 10^{-2}$, $\kappa_i = 10^{-5}$, $\mu_i = 0.003$ and $\chi_s = 0$. For the low-Mach number shock (Figure~\ref{F1}), ${\cal M}_0 = 1.423025$, $\Gamma_{ei} = 6\times 10^3$ and $\chi_a = 10^2$. For the high-Mach number shock (Figure~\ref{F2}), ${\cal M}_0 = 3.320392$, $\Gamma_{ei} = 10^7$ and $\chi_a = 10^3$. {\color{black}These parameters were chosen to produce results that exhibit three temperature behavior.} The spatial variable in Figures~\ref{F1} and \ref{F2} has been normalized to the radiation diffusion length scale in the post shock fluid, $L_r \equiv c/(3 v_1 \chi_a)$, where $v_1$ is the post-shock velocity in the shock frame.

It can be seen in Figures~\ref{F1} and \ref{F2} that separate ion and electron temperature spikes appear behind the shock, and that the precursor temperatures differ as well. The ions are directly heated by the compression, with the electrons being heated indirectly through their coupling to the ions. Conversely, the electrons are directly heated by the radiation, with the ions being indirectly heated. For the low-Mach number shock, the preferential heating of the electrons by the radiation in the precursor region results in an electron temperature spike that is larger than the ion temperature spike. For the high-Mach number shock, the electrons and ions are well-coupled in the precursor region, and the preferential response of the heavier ions to the compression results in that case in a larger ion temperature spike. It should be emphasized that even though a very narrow viscous layer can be seen in Figure~\ref{F1}, the solution is continuous.

Figure~\ref{F1} also shows results from \texttt{Kull}, a three temperature Lagrangian hydrodynamics code \cite{rath00}. The semi-analytic results from the relaxation code were imported onto a \texttt{Kull} mesh with $1000$ grid points, and the code was run for the time it took the shock to propagate across the computational domain. On this time scale, \texttt{Kull} quickly reaches a different steady state solution if there are any discrepancies between the numerical and semi-analytic results. Figure~\ref{F1} shows a good match between the two, and demonstrates the usefulness of shock tube problems for comprehensive coupled physics verification.

\begin{figure}
\includegraphics[scale=0.45]{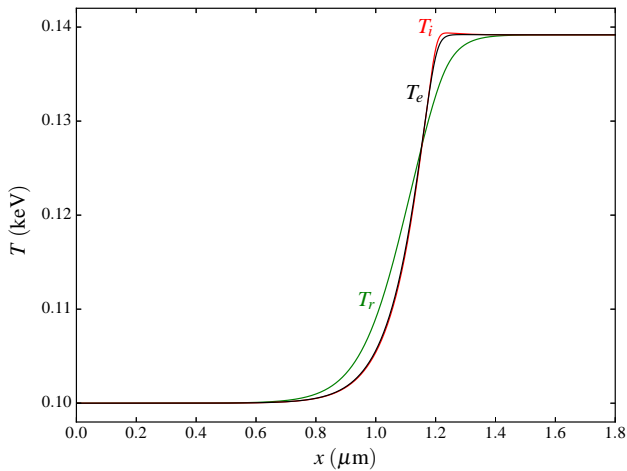}
\caption{\label{F3}Electron (\emph{black}), ion (\emph{red}) and radiation (\emph{green}) temperature profiles for a radiative plasma shock in the optically thin regime.}
\end{figure}

\begin{figure}
\includegraphics[scale=0.45]{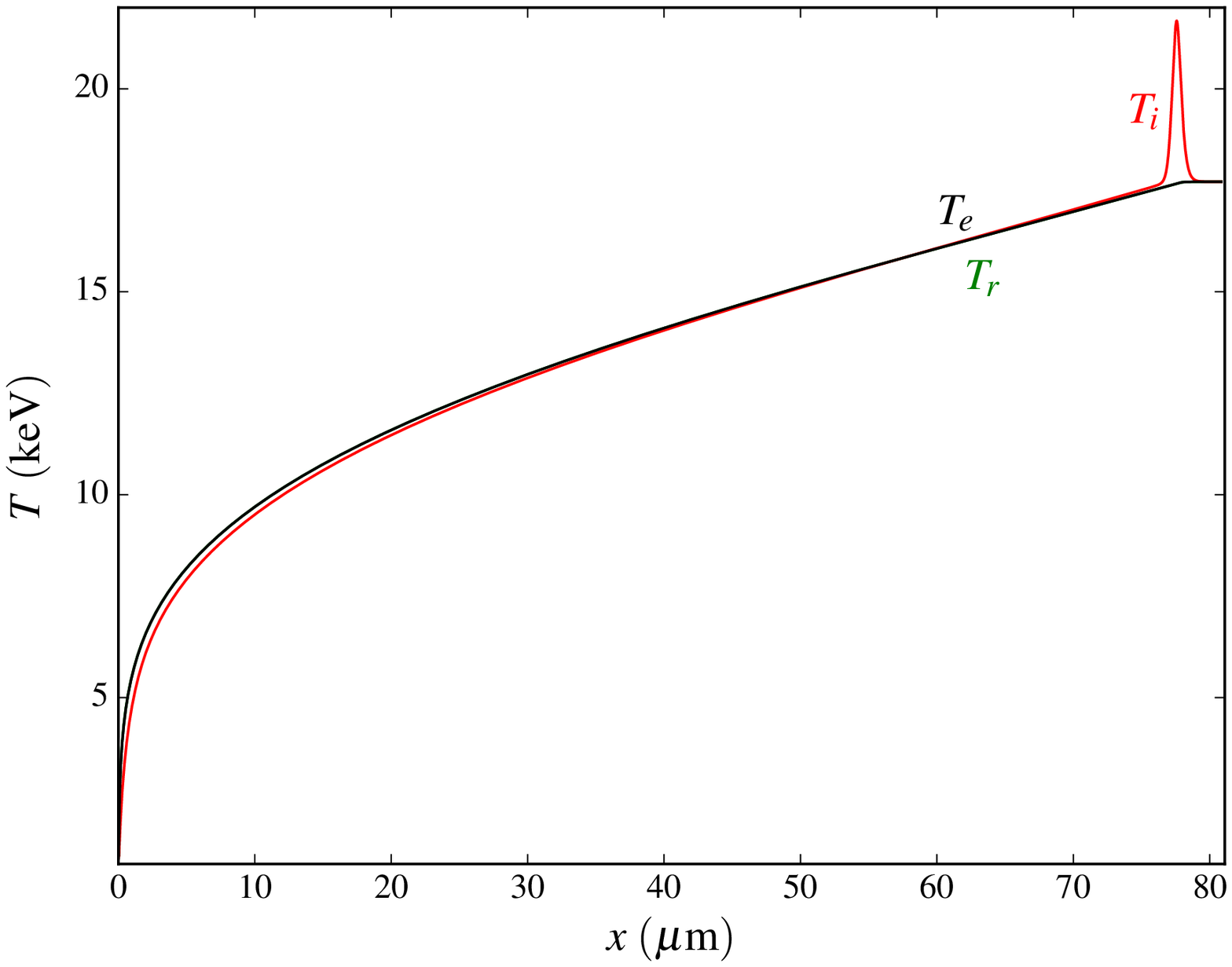}
\caption{\label{F4}Electron (\emph{black}), ion (\emph{red}) and radiation (\emph{green}) temperature profiles for a radiative plasma shock in the static diffusion regime.}
\end{figure}

\begin{figure}
\includegraphics[scale=0.45]{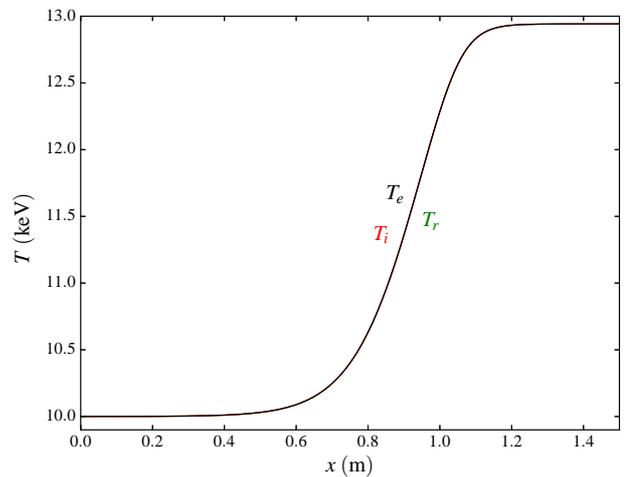}
\caption{\label{F5}Electron (\emph{black}), ion (\emph{red}) and radiation (\emph{green}) temperature profiles for a radiative plasma shock in the dynamic diffusion regime.}
\end{figure}

A wide variety of behavior can be obtained by varying the parameters in these solutions with constant coefficients. Rather than explore this unphysical parameter space, however, we proceed to representative solutions using physical models for the coefficients. For the conductivity, viscosity, and electron-ion coupling, we use the models in \cite{jp64}:
\[
\mu_i = \frac{5}{6}\sqrt{\frac{m_i}{\pi}}\frac{\left(k_B T_i\right)^{5/2}}{q^4 \ln \Lambda}, \;\; \kappa_i = \sqrt{\frac{m_e}{m_i}}\kappa_e = \frac{45 k_B \mu_i}{16 m_i},
\]
\[
\Gamma_{ei} = \frac{16 \sqrt{\pi} \rho^2 k_B q^4 \ln \Lambda}{m_e m_i^3 \left(2 k_B \left[T_e/m_e + T_i/m_i\right]\right)^{3/2}},
\]
where $q$ is the electron charge, $m_e$ is the electron mass, $\ln \Lambda$ is the Coulomb logarithm, and these expressions are valid for an atomic number $Z = 1$. We use $\ln \Lambda = 10$ for simplicity.

For the opacity, we use the bound-free and free-free expression of \cite{bl94} for the absorption, along with Thomson scattering:
\[
\chi_a = 1.5 \times 10^{20} \rho^2 T^{-5/2} \cm^{-1}, \;\; \chi_s = 0.348 \rho \cm^{-1},
\]
where in these expressions $\rho$ is in $\gcm$ and $T$ is in Kelvin.

Figures~\ref{F3}--\ref{F5} show results in the three regimes of radiation hydrodynamics: optically thin, static diffusion and dynamic diffusion \cite{mm84}. In the optically thin regime, the radiation is weakly coupled to the material. In the static diffusion limit, the radiation is thermally coupled to the material: radiation heats the material both before and after the shock front. The diffusive flux of radiation is important in this regime but radiation pressure is not (the jump conditions are hydrodynamic). In the dynamic diffusion limit, radiation pressure contributes significantly (the jump conditions are modified from the hydrodynamic case), and the radiation is both thermally and dynamically coupled to the material. 

\begin{figure}
\includegraphics[scale=0.45]{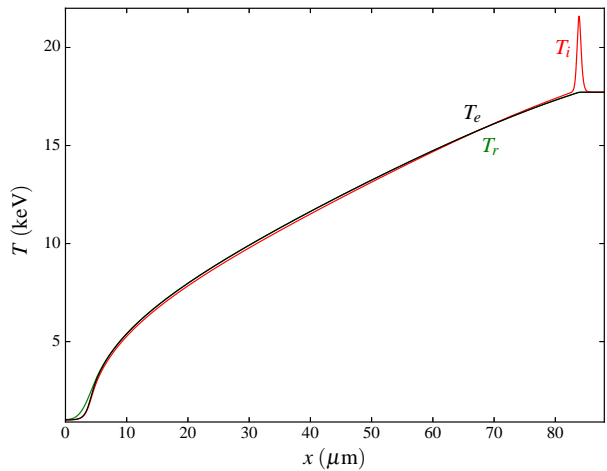}
\caption{\label{F6}Electron (\emph{black}), ion (\emph{red}) and radiation (\emph{green}) temperature profiles for a radiative plasma shock in the static diffusion regime with $\chi_a = 10^4 \cm^{-1}$.}
\end{figure}

\begin{figure}
\includegraphics[scale=0.45]{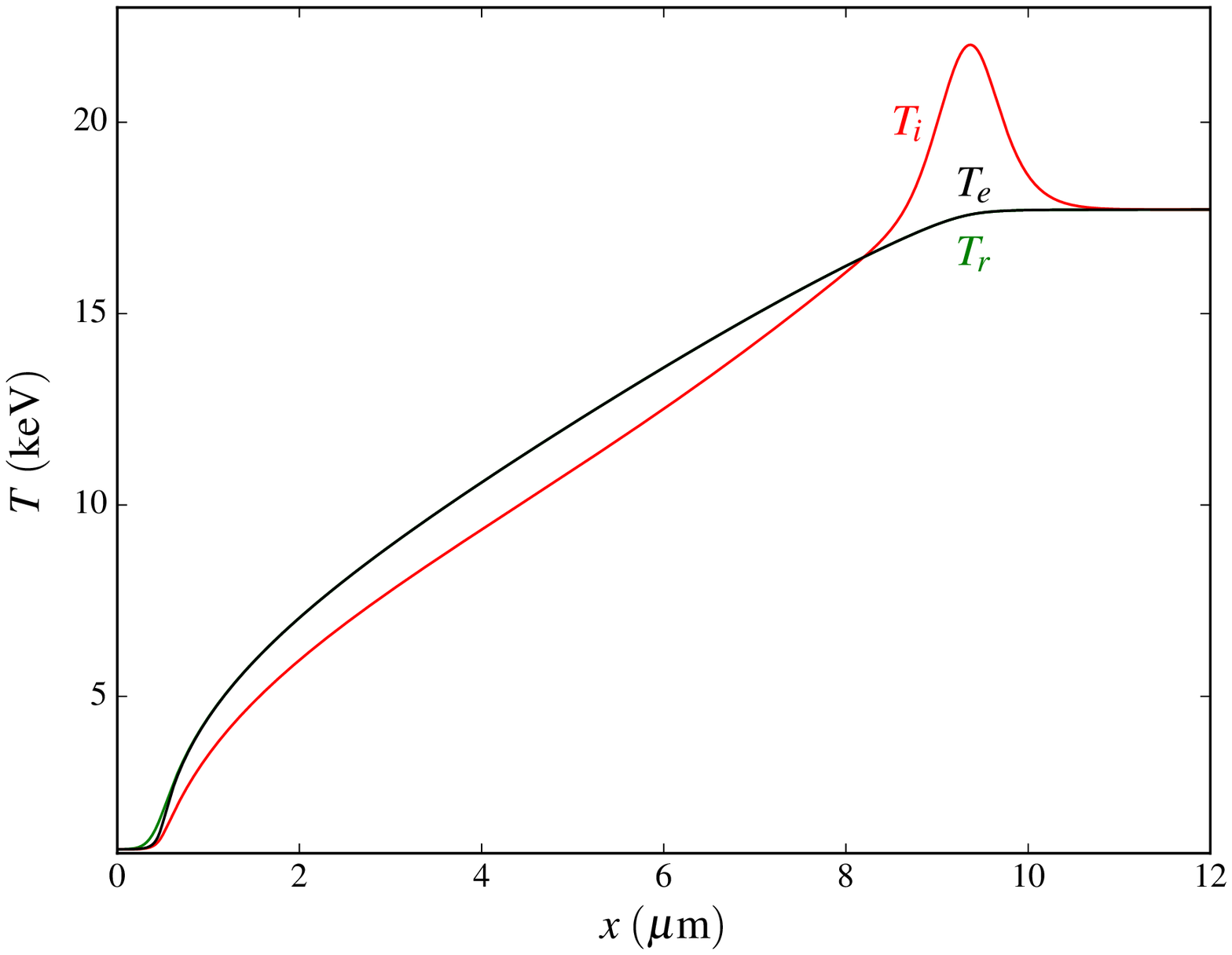}
\caption{\label{F7}Electron (\emph{black}), ion (\emph{red}) and radiation (\emph{green}) temperature profiles for a radiative plasma shock in the static diffusion regime with $\chi_a = 10^5 \cm^{-1}$.}
\end{figure}

\begin{figure}
\includegraphics[scale=0.45]{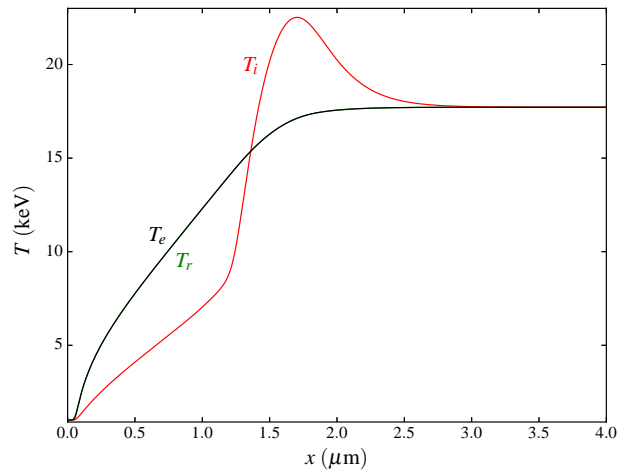}
\caption{\label{F8}Electron (\emph{black}), ion (\emph{red}) and radiation (\emph{green}) temperature profiles for a radiative plasma shock in the static diffusion regime with $\chi_a = 10^6 \cm^{-1}$.}
\end{figure}

The common parameters used for these calculations were $\gamma = 5/3$ and $C_{ve} = C_{vi} = 0.072364$. For the optically thin calculation (Figure~\ref{F3}), the additional parameters used were $\rho_0 = 1 \gcm$, $T_0 = 0.1 \kev$ (corresponding to ${\cal C}_0 = 2364$, ${\cal P}_0 = 8.532 \times 10^{-5}$) and ${\cal M}_0 = 1.4$. All three components have distinct temperatures in this case.\footnote{{\color{black}This case is actually marginally optically thin (there are a few optical depths across the shock); solutions are difficult to calculate as the material becomes more optically thin due to the large disparity between the shock width and the length scale associated with the free streaming radiation.}} For the static diffusion calculation (Figure~\ref{F4}), the parameters used were $\rho_0 = 100 \gcm$, $T_0 = 1 \kev$ (corresponding to ${\cal C}_0 = 747.6$, ${\cal P}_0 = 8.532 \times 10^{-4}$) and ${\cal M}_0 = 10$. At these temperatures, the opacity is sufficiently large that the electron and radiation temperatures are nearly equivalent, resulting in two-temperature behavior. {\color{black}The optical depth across the shock in Figure~\ref{F4} satisfies $\beta \tau \sim 0.3 \ll 1$, where $\beta = v/c$ and $\tau = \int \left(\chi_a + \chi_s\right) \, dx$.} For the dynamic diffusion calculation (Figure~\ref{F5}), the parameters used were $\rho_0 = 1 \gcm$, $T_0 = 10 \kev$ (corresponding to ${\cal C}_0 = 236.4$, ${\cal P}_0 = 85.32$) and ${\cal M}_0 = 10$. All three components are well coupled in this case and one-temperature behavior occurs.  {\color{black}The optical depth across the shock in Figure~\ref{F5} satisfies $\beta \tau \sim 7 \gg 1$.} 

Figure~\ref{F4} is qualitatively different from previous plasma shock solutions \cite{zr67,mwl11}. This is due to the fact that previous studies have ignored radiation, which heats the electrons on length scales that are much longer than the electron conduction and Coulomb coupling length scales. Figures~\ref{F6}--\ref{F8} show results from a series of calculations with the same parameters as in Figure~\ref{F4}, using the physical conduction and electron-ion coupling models but a constant opacity of $\chi_a = 10^4$, $10^5$ and $10^6 \cm^{-1}$. The material is sufficiently opaque in all four cases to keep the electrons and radiation at the same temperature (i.e., the mean free path of the photons is short compared to the shock width), but the two-temperature results are qualitatively different depending on the value of the opacity. As the opacity increases, the radiation diffusion length scale decreases. In Figure~\ref{F6}, which most closely matches the physical case (Figure~\ref{F4}), the radiation diffusion length scale is much longer than the Coulomb coupling length scale, whereas in Figure~\ref{F8} it is much shorter. {\color{black}Previous plasma shock solutions, which resemble the results in Figure~\ref{F8}, are valid at sufficiently low temperatures that the radiation flux is negligible.}

\section{\label{DISCUSSION}Discussion}

We have obtained for the first time steady-state solutions for three-temperature shocks. These solutions represent the most comprehensive multi-physics verification test developed to date. While we are not the first to use shock tubes for code verification \cite{le08}, we want to emphasize their utility in that regard. Even though solutions of this type are only valid for planar geometry and a steady state, no approximations have been applied to the various terms in the original equation set on which our analysis is based. The methodology is thus competitive with the Method of Manufactured Solutions (MMS) as far as code coverage is concerned, and provides two additional benefits at the same time: 1) the incorporation of more complicated physics models is trivial, whereas the complexity of MMS increases dramatically with additional complexity, and 2) physical intuition is a guide rather than a hindrance. We believe that the relaxation-based approach we have used here is {\color{black}the best avenue} for further multi-physics verification developments along these lines.

We have obtained results in all three regimes of radiation hydrodynamics: optically thin, static diffusion, and dynamic diffusion (Figures~\ref{F3}--\ref{F5}). As far as we are aware, the latter is the first result of its kind. Due to its high velocity and large shock width, the dynamic diffusion shock solution is likely relevant only in an astrophysical context {\color{black}(e.g., radiation dominated accretion flows associated with black holes or accretion onto the surface of a magnetized neutron star)}. The primary physical conclusion from this work is that radiation has a non-negligible impact on plasma shocks at sufficiently high temperatures. Radiation heats the electrons, which in turn heat the ions via Coulomb collisions, and this heating takes place on diffusive length scales that are much longer than the length scale associated with electron conduction and Coulomb coupling. As a result, our plasma shock solutions resemble radiative shock solutions \cite{le08} (with the electrons coupled to the radiation) rather than plasma shock solutions that neglect radiation \cite{zr67,mwl11}.

\begin{acknowledgements}

We thank Jim Ferguson, Miguel Holgado, Rob Lowrie, George Zimmerman and the referees for helpful discussions and comments. This work was performed under the auspices of Lawrence Livermore National Security, LLC, (LLNS) under Contract No.$\;$DE-AC52-07NA27344.

\end{acknowledgements}

\begin{appendix}

\section{Details of solution methodology}\label{APP}

{\color{black}We first demonstrate the presence of non-shock solutions in a reduced version of our system of equations.} Neglecting ion conduction and radiation, equations (\ref{INT1}) and (\ref{INT2}) are equivalent to the equations for a hydrodynamic shock with viscosity and heat conduction (for arbitrary Prandtl number $\Pran \equiv \mu_i \gamma C_v/\kappa_e$). For $\Pran = 3/4$, equation (\ref{INT2}) can be integrated to give \cite{bec22,joh13}
\be\label{INT3}
\frac{v^2}{2} + h = c_2 + A \, e^{x/L_{e}},
\ee
where $L_{e} \equiv \kappa_e/(\gamma C_v m_0)$ and $A$ is an arbitrary constant. Satisfying the shock boundary conditions as $x \to \infty$ requires $A = 0$, although the above is the general expression obeying momentum and energy conservation. The non-shock portion can be removed by hand by setting $A = 0$, but this requires first obtaining the integral (\ref{INT3}). For a numerical integration of equations (\ref{INT1}) and (\ref{INT2}), it is necessary to integrate from the post-shock region so that the exponential in (\ref{INT3}) decays as the integration advances.

Even with $A = 0$, non-shock solutions exist. For example, for $\Pran = 3/4$, $A = 0$ and $M_0^2 = 4/(3-\gamma)$, equations (\ref{INT1}) and (\ref{INT2}) have the closed-form solutions \cite{joh14}
\be\label{R2}
\frac{v}{v_0} = 1 + \frac{f}{2} \pm \sqrt{\left(\frac{f}{2}\right)^2 + \frac{f}{2}}, \;\;
f \equiv \exp\left(\frac{\gamma+1}{2\gamma}\frac{x}{L_{e}}\right).
\ee
The negative branch above is associated with the shock solution, whereas the positive branch is associated with a solution that grows without bound as $x \to \infty$; the temperature decreases from its ambient value in this solution and eventually goes negative. While this non-physical behavior is perhaps surprising, it is possible that this solution is unstable and will therefore not be an attractor in a time-dependent calculation, or that the exponential in (\ref{INT3}) acts to prevent negative temperatures from occurring. Up until the point at which the temperature goes negative, the non-shock solution associated with the positive branch of (\ref{R2}) can be regarded as a planar wind expanding into a cold vacuum.

\begin{figure}
\includegraphics[scale=0.45]{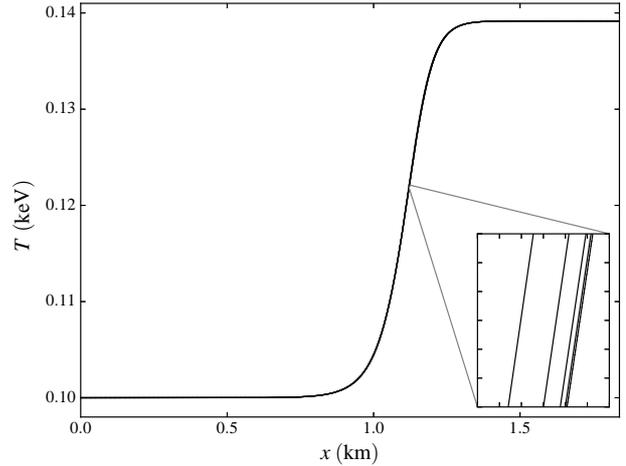}
\caption{\label{F9}{\color{black}Convergence of a $\Pran = 10^5$ numerical solution ($N = 10^3$, $3\times 10^3$, $10^4$, $3\times 10^4$ and $10^5$ from left to right). The inset is $150\times$ magnification.}}
\end{figure}

\begin{figure}
\includegraphics[scale=0.45]{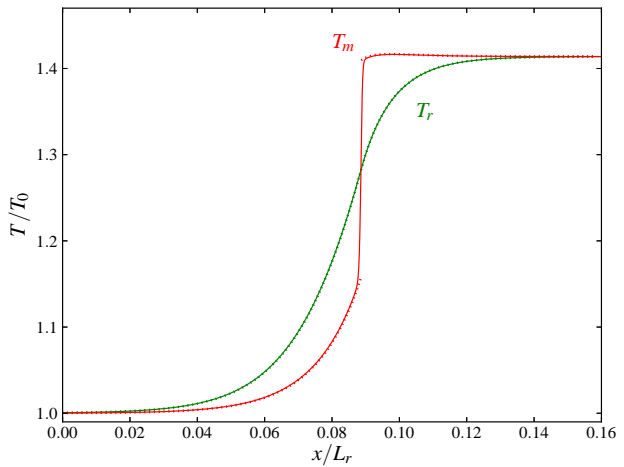}
\caption{\label{F10}Material (\emph{red}) and radiation (\emph{green}) temperature profiles for a shock with the same parameters as Figure~\ref{F1} except $\Gamma_{ei} = 10^8$, using a three-temperature (\emph{solid}) and a two-temperature (\emph{dotted}) solver. The latter solution is discontinuous.}
\end{figure}

The full set of equations to be solved are equations (\ref{MOM})--(\ref{ER}) plus the four equations associated with the flux definitions, for a total of eight equations. {\color{black}The eight unknowns are the three temperatures, the four fluxes, and the velocity. We have experimented with various methods of solving these equations. For the solutions with constant coefficients (Figures~\ref{F1}--\ref{F2} and \ref{F10}), we solve them directly (using $x$ as the independent variable) with the relaxation algorithm \texttt{solvde} of \cite{pre92}. We have found it useful in that case to add an arbitrary scaling factor to the spatial variable that we can adjust on the fly, since the width of the solution can change by orders of magnitude as we vary parameters. For the solutions shown in Figures~\ref{F3}--\ref{F9}, we use} $v$ rather than $x$ as the independent variable, solving the following seventh-order system of equations with the relaxation algorithm:
\be\label{EQ1}
\left(\dv{v}{x}\right)\dv{T_\alpha}{v} = -\frac{F_\alpha}{\kappa_\alpha},
\ee
\be
\left(\dv{v}{x}\right)\dv{F_e}{v} = \frac{m_0 C_{ve}}{\kappa_e} F_e - p_e \dv{v}{x} - S_{ei} - S_{er},
\ee
\be
\left(\dv{v}{x}\right)\dv{F_i}{v} = \frac{m_0 C_{vi}}{\kappa_i} F_i - p_i \dv{v}{x} + S_{ei} + \frac{4\mu_i}{3}\left(\dv{v}{x}\right)^2,
\ee
\be
\left(\dv{v}{x}\right)\dv{F_r}{v} = \frac{3 \chi_t v}{c} F_r - 4 p_r \dv{v}{x} + S_{er},
\ee
\bea\label{EQ7}
\left(\dv{v}{x}\right)\dv{F_{v}}{v} &=& \left(\frac{p_e + p_i}{v} - m_0\right)\dv{v}{x} \nonumber \\
&+& (\gamma - 1) \rho \left(\frac{C_{ve}}{\kappa_e}F_e + \frac{C_{vi}}{\kappa_i}F_i\right) + \frac{\chi_t}{c}F_r,
\eea
where equation (\ref{EQ7}) was obtained by expanding $dp/dx$ in equation (\ref{MOM}), {\color{black}$\rho = m_0/v$, and $p_{e,i}$ are given by the equations of state. The eighth equation in this case is} 
\be\label{EQ8}
\dv{x}{v} = -\frac{4\mu_i}{3 F_{v}},
\ee
{\color{black}which we solve by simple quadrature to obtain $x(v)$.}\footnote{{\color{black}Even though the velocity flux can be very small, it is never zero, and we have not encountered any difficulties with integrating equation (\ref{EQ8}) directly without approximation.}} {\color{black}To obtain the shock jump conditions, we use the globally convergent Newton's method algorithm \texttt{newt} from \cite{pre92}. We obtain the same overall jump conditions as \cite{le08}.}

One advantage of relaxation algorithms is that they naturally handle singularities. Both the right hand side and the factor $dv/dx$ in equations (\ref{EQ1})--(\ref{EQ7}) go to zero at the end points, and writing the equations in this way avoids the $0/0$ situation encountered in a straightforward integration from the end points \cite{mwl11,le08}. A guess for the solution is initialized on a grid of $N$ points in velocity space, where we use a separate logarithmic spacing in the pre- and post-shock regions in order to avoid poor resolution as $|x| \to \infty$:
\[
v_k =
\left\{
\begin{array}{cc}
v_0 - \epsilon_- v_0 \left(\frac{v_0 - \overline{v}}{\epsilon_- v_0}\right)^\frac{k}{N/2-1} & \; {\rm for} \; k < N/2 \\
v_1 + \epsilon_+ v_1 \left(\frac{\overline{v} - v_1 + \delta v}{\epsilon_+ v_1}\right)^\frac{N-k}{N/2} & \; {\rm for} \; k \geq N/2, \\
\end{array}
\right.
\]
where $\overline{v} \equiv \sqrt{v_0 v_1}$,
\[
\delta v \equiv \overline{v} - v_0 + \epsilon_- v_0 \left(\frac{v_0 - \overline{v}}{\epsilon_- v_0}\right)^\frac{N/2-3}{N/2-2},
\]
and $\epsilon_\pm$ are small numbers that determine how far the solution extends into the pre- and post-shock equilibrium regions.

In the absence of viscosity, some of the shock solutions are discontinuous, and a fair amount of analysis would need to go into determining whether a discontinuity is present and how to handle one when present. In addition, it is not clear that a relaxation algorithm could handle a discontinuity in the middle of its solution domain; it may be possible to obtain two relaxation solutions on either side of the discontinuity, but determining the boundary conditions to apply at the discontinuity would be difficult. Including viscosity in the set of equations to be solved avoids all of these issues and allows for a single solution to be determined across the domain. {\color{black}At the same time, our approach has the disadvantage of being unable to address the question of whether or not a discontinuity is present in the absence of viscosity.}

{\color{black}We have generally found that $N \sim 10^4$ gives a sufficiently converged result (i.e., a result that is insensitive to $N$; see Figure~\ref{F9}). For solutions that have an embedded viscous layer, the resolution requirement depends upon the width of the viscous layer and therefore the precise value of the viscosity. We have found that the relaxation algorithm becomes numerically unstable unless there are at least $\sim 10$ points across the viscous layer. The solutions shown in Figure~\ref{F1} have $\sim 30$ points across the viscous layer and therefore easily satisfy this requirement.}

Another advantage of including viscosity is that it allows for a simple analytical solution to be used as the initial guess for the relaxation algorithm. In the large $\Pran$ limit (viscosity dominating conductivity), the thermal fluxes can be ignored, and an analytical solution to equations (\ref{INT1}) and (\ref{INT2}) can be derived as in \cite{joh13}. For $F_{e} = F_i = 0$, equations (\ref{INT1}) and (\ref{INT2}) can be combined to give the following quartic equation for $T_k(v_k)$:
\be\label{QUARTIC}
a_1 T_k^4 + a_2 T_k + a_3 = 0,
\ee
where
\[
a_1 \equiv \frac{a_r v_k}{m_0} ,\;\; a_2 \equiv C_{v} , \;\; a_3 \equiv -\frac{1}{2} v_k^2 + c_1 v_k - c_2.
\]
The solution to (\ref{QUARTIC}) appropriate for a shock is
\be\label{TI}
T_k = -S + \sqrt{\frac{a_2}{4a_1S} - S^2},
\ee
where
\[
S \equiv \frac{1}{2}\sqrt{\frac{Q + D_0/Q}{3a_1}},\;\; Q \equiv \left(\frac{D_1 + \sqrt{27\Delta}}{2}\right)^{1/3},
\]
\[
D_0 \equiv 12 a_1 a_3 , \;\; D_1 \equiv 27 a_1 a_2^2, \;\; \Delta \equiv 27 a_1^2 a_2^4 - 256 a_1^3 a_3^3.
\]

If the electron-ion coupling coefficient and opacity are set sufficiently large, equation (\ref{TI}) can be used to initialize all three temperatures. The velocity flux can be determined from equation (\ref{INT1}):
\be\label{FVI}
F_{v,k} = m_0 c_1 - m_0 v_k - (\gamma-1)m_0 C_v \frac{T_k}{v_k} - \frac{1}{3} a_r T_k^4,
\ee
and the thermal fluxes can be expressed in terms of $F_{v,k}$ by differentiating equation (\ref{QUARTIC}) with respect to $v_k$:
\be\label{FTI}
F_{\alpha,k} = \left(\frac{3\kappa_\alpha}{4\mu_i} \right)\left(\frac{v_k - c_1 - a_r T_k^4/m_0}{a_2 + 4 a_1 T_k^3}\right) F_{v,k}.
\ee
Notice that the other three solutions to the quartic equation (\ref{QUARTIC}) constitute additional non-shock solutions that are potential attractors for a numerical solver. {\color{black}The spatial variable can be obtained by numerically integrating equation (\ref{EQ8}), or, when radiation pressure is negligible, by using the expression for $x(v)$ derived in \cite{joh13}.} 

{\color{black}Figure~\ref{F9} shows the solution obtained from the relaxation algorithm using the analytical solution described above as the initial guess, with $\rho_0 = 1 \gcm$, $T_0 = 0.1 \kev$ and $M_0 = 1.4$ (as in Figure~\ref{F3}), constant $\Gamma_{ei} = \chi_a = 10^6$ and $\Pran = 10^5$. The highest resolution numerical solution in Figure~\ref{F9} is indistinguishable from the analytical solution.} In practice, we have found that $\Pran = 100$ suffices to obtain an initial solution. Having obtained this solution, the next step is to vary the coefficients to their desired values, stepping slowly through parameter space. We transition from constant to physical coefficients by introducing a numerical parameter into each coefficient that allows for a smooth transition between the two states. The viscosity, for example, is set to
\[
\mu_i = \mu_{ic}\left(\frac{\mu_{ip}}{\mu_{ic}}\right)^\eta,
\]
where $\mu_{ic}$ is the constant coefficient, $\mu_{ip}$ is the physical model, and we vary $\eta$ slowly from $0$ to $1$. {\color{black}The material parameters ($\mu_i$, $\kappa_e$, $\kappa_i$ and $\Gamma_{ei}$) are varied simultaneously in the same manner, with the same parameter $\eta$.} For the results shown in Figures~\ref{F3}--\ref{F5}, we increased $\eta$ in increments of $0.01$. The opacity was sufficiently large in these calculations that we were able to generate these results by using the physical models for the opacity from the start.

Additional practical considerations for this solution methodology are as follows:
\begin{itemize}
\item While the number of equations to be solved can be reduced by using the integrals (\ref{INT1}) and (\ref{INT2}), we have found solving the full set of equations to be more robust.
\item {\color{black}We have experimented with various boundary conditions, and have found that} applying boundary conditions to the fluxes is more robust than applying boundary conditions to the temperatures.
\item {\color{black}For the results shown in Figures~\ref{F1}--\ref{F2} and \ref{F10}, we applied boundary conditions to the velocity and material temperatures at both ends of the computational domain, along with the radiation temperature at the far end of the shocked fluid.
\item For the results shown in Figures~\ref{F3}--\ref{F8}, we applied boundary conditions to the thermal fluxes at both ends of the computational domain, along with the velocity flux at the far end of the shocked fluid.
\item We generally set the boundary fluxes equal to values from the initial analytical guess (these are small but nonzero). Often, however, the fluxes can change by orders of magnitude (even though they remain small), and this can generate numerical instability at the boundary. We have also experimented with extrapolating the fluxes from the interior of the domain out to the boundaries, and this appears to be a more robust approach.}\footnote{{\color{black}We extrapolate from a point that is away from the boundaries but still within the region in which departures from equilibrium are small.}}
\item Moving around in $\rho_0$, $T_0$, ${\cal M}_0$ space is difficult and usually results in numerical instability. As a result, we have found it necessary to keep the jump conditions fixed for a given solution and simply change the spatial profile across the shock.
\item The fluxes can vary by many orders of magnitude between the end points and the shock front, and we have found it necessary in most cases to resort to \texttt{long double} precision in our \texttt{C} implementation.
\item A useful diagnostic is to monitor the fluxes as the solution proceeds; a failure of the relaxation algorithm is usually associated with noise in the fluxes near the end points of the solution.
\item {\color{black}Analytical solutions also exist for $\Pran = 3/4$ and could be used as an initial guess \cite{joh13,joh14}.}
\item {\color{black}Setting boundary conditions based upon a linear eigenvalue analysis near the end points may provide a more robust solution methodology \cite{juk57}.}
\end{itemize}

As an additional verification of our relaxation algorithm, Figure~\ref{F10} shows a comparison between a calculation using it and a calculation using the standard approach (integrating from the end points to the middle and matching with an embedded hydrodynamic shock). The parameters used for this calculation were the same as those in Figure~\ref{F1} with the electron-ion coupling coefficient set to a large value ($\Gamma_{ei} = 10^8$) in order to fix the electrons and ions at the same temperature. This enables a direct comparison with a two-temperature radiative shock solution, {\color{black}which we obtain with a \texttt{scipy} integrator using the methodology described in} \cite{le08}.

\end{appendix}


\begin{thebibliography}{}
\bibitem{zr67} Zel'dovich, Y.~B., Raizer, Y.~P.: Physics of shock waves and high-temperature hydrodynamic phenomena. Academic Press, New York (1966)  

\bibitem{jp64} Jaffrin, M.~Y., Probstein, R.~F.: Structure of a plasma shock wave. Phys. Fluids {\bf 7}, 1658--1674 (1964)

\bibitem{mwl11} Masser, T.~O., Wohlbier, J.~G., Lowrie, R.~B.: Shock wave structure for a fully ionized plasma. Shock Waves {\bf 21}, 367--381 (2011)

\bibitem{mm84} Mihalas, D., Mihalas, B.~W.: Foundations of radiation hydrodynamics. Oxford University Press, New York (1984)  

\bibitem{le08} Lowrie, R.~B., Edwards, J.~D.: Radiative shock solutions with grey nonequilibrium diffusion. Shock Waves {\bf 18}, 129--143 (2008)

{\color{black}\bibitem{mk82} Mihalas, D., Klein, R.~I.: On the solution of the time-dependent inertial-frame equation of radiative transfer in moving media to O(v/c). J. Comp. Phys. {\bf 46}, 97--137 (1982)

\bibitem{nem60} Nemytskii, V., Stepanov, V.: Qualitative theory of differential equations. Princeton University Press, Princeton (1960)  

\bibitem{and63} Anderson, J.~E.: Magnetohydrodynamic shock waves. MIT Press, Cambridge (1963)  

\bibitem{gro65} Gross, R.~A.: Strong ionizing shock waves. Rev. Mod. Phys. {\bf 37}, 724--743 (1965)}

\bibitem{juk57} Jukes, J.~D.: The structure of a shock wave in a fully ionized gas. J. Fluid Mech. {\bf 3}, 275--285 (1957)

\bibitem{bec22} Becker, R.: Stosswelle und Detonation. Z.~Physik. {\bf 8}, 321--362 (1922)

\bibitem{joh13} Johnson, B.~M.: Analytical shock solutions at large and small Prandtl number. J. Fluid Mech. {\bf 726}, R4 (2013)

\bibitem{joh14} Johnson, B.~M.: Closed-form shock solutions. J. Fluid Mech. {\bf 745}, R1 (2014).

\bibitem{pre92} Press, W.~H., Teukolsky, S.~A., Vetterling, W.~T., Flannery B.~P.: Numerical Recipes in C. University Press, Cambridge (1992)  

\bibitem{rath00} Rathkopf, J.~A., Miller, D.~S., Owen, J.~M., Stuart, L.~M., Zika, M.~R., Eltgroth, P.~G., Madsen, N.~K., McCandless, K.~P., Nowak, P.~F., Nemanic, M.~K., Gentile, N.~A., Keen, N.~D. KULL: LLNL's ASCI inertial confinement fusion simulation code. Physor 2000, ANS Int. Topical Mtg. Adv. in Reactor Phys. \& Math. \& Comput. into the Next Millennium (2000)

\bibitem{bl94} Bell, K.~R., Lin, D.~N.~C.: Using FU Orionis outbursts to constrain self-regulated protostellar disk models. Astrophys. J. {\bf 427}, 987--1004 (1994)

\end{thebibliography}


\end{document}